\documentclass[fleqn,usenatbib]{mnras}

\usepackage{newtxtext,newtxmath}

\usepackage[T1]{fontenc}

\DeclareRobustCommand{\VAN}[3]{#2}
\let\VANthebibliography\thebibliography
\def\thebibliography{\DeclareRobustCommand{\VAN}[3]{##3}\VANthebibliography}

\usepackage{graphicx}	
\usepackage{amsmath}	
\usepackage{float}
\usepackage{lipsum}
\usepackage{multirow}
\usepackage{xcolor}

\title[Measurement of galaxy clustering and galaxy bias]{Galaxy clustering measurements out to redshift  z$\sim$8 from Hubble Legacy Fields}

\author[Nicolò Dalmasso et al.]{
Nicolò Dalmasso$^{1,2}$\thanks{e-mail: ndalmasso@student.unimelb.edu.au},
Michele Trenti$^{1,2}$,
Nicha Leethochawalit$^{3}$
\\
$^{1}$School of Physics, University of Melbourne, Parkville, Vic 3010, Australia\\
$^{2}$ARC Centre of Excellence for All Sky Astrophysics in 3 Dimensions (ASTRO 3D), Australia\\
$^{3}$National Astronomical Research Institute of Thailand (NARIT), Mae Rim, Chiang Mai, 50180, Thailand\\
}

\date{Accepted XXX. Received YYY; in original form ZZZ}

\pubyear{2023}

\begin{document}
\label{firstpage}
\pagerange{\pageref{firstpage}--\pageref{lastpage}}
\maketitle

\begin{abstract}
We present a novel approach for measuring the two-point correlation function of galaxies in narrow pencil beam surveys with varying depths. Our methodology is utilized to expand high-redshift galaxy clustering investigations up to $z \sim 8$ by analyzing a comprehensive sample consisting of $N_g = 160$ Lyman break galaxy candidates obtained through optical and near-infrared photometric data within the CANDELS GOODS datasets from the Hubble Space Telescope Legacy Fields. For bright sources with $M_{UV} < -19.8$, we determine a galaxy bias of $b = 9.33\pm4.90$ at $\overline{z} = 7.7$ and a correlation length of $r_0 = 10.74\pm7.06$ $h^{-1}Mpc$. We obtain similar results for the XDF, with a galaxy bias measurement of $b = 8.26\pm3.41$ at the same redshift for a slightly fainter sample with a median luminosity of $M_{UV} = -18.4$. By comparing with dark-matter halo bias and employing abundance matching, we deduce a characteristic halo mass of $M_h \sim 10^{11.5} M_{\odot}$ and a duty cycle close to unity. To validate our approach for variable-depth datasets, we replicate the analysis in a region with near-uniform depth using a standard two-point correlation function estimator, yielding consistent outcomes. Our study not only provides a valuable tool for future utilization in JWST datasets but also suggests that the clustering of early galaxies continues to increase with redshift beyond $z \gtrsim 8$, potentially contributing to the existence of protocluster structures observed in early JWST imaging and spectroscopic surveys at $z \gtrsim 8$.

\end{abstract}




 \section{Introduction}\label{intro}
\indent Studying galaxy clustering, i.e. quantifying the non-uniform angular distribution of galaxies in the sky, is a fundamental tool to investigate how the structures we observe today formed and evolved during the course of the history of the Universe (e.g.,\citealt{1996MNRAS.282..347M}). In particular, clustering measurements at high redshift offer insight to link the rapid build-up of the galaxy luminosity function (LF) to the assembly of dark-matter halos (e.g., \citealt{2005MNRAS.364..303C,2006astro.ph..1090C,Lee_2006,2008MNRAS.383..355V}).\\
\indent A complementary approach to investigate this is through abundance matching \citep{10.1111/j.1365-2966.2004.08059.x}, this technique involves comparing the number of galaxies with luminosity exceeding a specified threshold $L_{g}$ to the amount of halos with masses surpassing a given threshold $M_h$ under two assumptions: each halo or sub-halo is hosting an individual galaxy and the luminosity and the halo mass functions are monotonically related (e.g.,\citealt{martinez_2002, neyrinck}). Through the combination of clustering measurements and abundance matching analysis, the galaxy duty cycle ($\epsilon_{DC}$) can be constrained \citep{martinez_2002}. Also referred as occupation fraction this parameter is defined as the fraction of UV-bright Lyman break galaxies (LBGs) that reside within a dark matter halo, and it is an important quantity to constrain average star formation efficiency within dark matter halos.\\
\indent Over the past two decades, clustering measurements have been performed out to $z\lesssim7$. Using data from Hubble Deep Field North, \citep{10.1046/j.1365-8711.1999.02978.x} quantified the redshift evolution of clustering between redshift $0\leq z\leq4.5$ for photometrically selected LBGs. The angular correlation function (ACF) derived showed a clear decrease in amplitude from $z=0$ up to a redshift $z\sim 1-1.5$ followed by a slow increase in amplitude at higher redshift, with the same trend found in the correlation length. These two behaviours translate into an increasing galaxy bias $b(z)$ at $z\gtrsim 1.5$ showing that a typical minimum halo mass of $M_h\gtrsim10^{12}h^{-1}M_{\odot}$ is needed for galaxies with  $b(z\simeq4)\simeq3$. Galaxy bias, denoted as $b$, represents the systematic deviation or clustering of galaxies compared to the underlying matter distribution in the universe. It serves as an indicator of the non-uniform spatial distribution of galaxies, influenced by dark matter distribution and the large-scale structure of the universe. An increasing trend in galaxy bias implies stronger galaxy clustering and deviation from the matter distribution in larger cosmic structures.\\ 
\indent As we move to higher redshifts, it has been deduced that the minimum dark matter halo mass must fall within the range of $M_{h}\sim 10^{11.0-11.5}M_{\odot}$ for galaxies situated at a redshift of approximately $z\simeq3-3.5$ \citep{Giavalisco_2001,adelberger_2005}. The increase of $b(z)$ with $z$ continues at $z\geq3.5$ and the characteristic halo mass remains close to $M_h\sim5\cdot10^{11}M_{\odot}$ at $z\simeq3.5-6$ \citep{Ouchi_2004,Overzier_2006}.\\ 
\indent Using LBGs candidates at $3.5\leq z\leq5.5$, \citet{Lee_2006} investigated a more realistic model for the ACF with a function composed of two terms, a correlation  between two halos on relatively large separations ($w_{2h}(\theta)$) and a one-halo term associated to presence of multiple galaxies in a single halo ($w_{1h}(\theta)$). The latter dominates the correlation signal at very small separations (comparable to the typical virial radius of a halo with $M_h\gtrsim 10^{12}~\mathrm{M_{\odot}}$).\\ 
\indent In their work, \citet{wyithe_2014} derived a theoretical estimate of $\epsilon_{DC}(z)\simeq 0.10-0.15$ for $z\gtrsim 6$ using a combination of clustering and abundance matching analyses. This approach aimed to reconcile the observed discrepancies in total stellar mass and associated star formation rates.\\ 
\indent This theoretical prediction has been subsequently tested by \citet{Barone_2014} with new galaxy clustering measurements out to $z=7.2$ using LBGs from the GOODS/CANDELS and XDF Hubble legacy fields observations \citep{Bouwens_2015}. Their analysis of the evolution of galaxy bias with redshift shows an increasing trend, reaching $b=9.7^{+2.0}_{-2.5}$ for their sample of UV bright LBGs at redshift $z=7.2$. This translate to a minimum dark matter halo mass of $M_h\gtrsim10^{11.0}M_{\odot}$.\\
\indent Using a comparable sample of LBGs derived from the Hubble Legacy deep imaging, \citet{Harikane_2016} conducted a study on the angular correlation function up to redshift $z=6.8$, focusing on UV bright candidates ($m_{UV}<27.6$). Through their measurements, they deduced a range of acceptable halo masses for galaxies residing within the redshift interval of $z\sim 6-7$ to be approximately $M_h\sim(1-20)\cdot 10^{11} M_{\odot}$, in accordance with what had been found previously in the high redshift regime ($z\sim5.8$) (e.g.,\citealt{Hamana_2004,Lee_2006,Hildebrandt_2009,Barone_2014}).\\
\indent In contrast to \citet{Harikane_2016}'s redshift $z=5.8$ findings of a galaxy bias of $b=4.7^{+1.0}_{-1.3}$, both \citet{Hatfield_2018} and \citet{Qiu_2018} report slightly higher galaxy bias values for galaxies with similar redshift and magnitude when estimating halo mass, measuring $b=6.0\pm1.1$ and $b=6.1^{+1.6}_{-2.0}$ respectively. These differences may be due to computation method variations, but clustering trends with luminosity remain consistent. Both studies find significantly higher typical host halo mass and galaxy bias in their samples compared to lower luminosity ones.\\
\indent In their subsequent study \citet{Harikane_2022}, the authors focus on measuring the rest-UV luminosity functions and angular correlation functions across a wide range of redshifts, specifically $z\sim 2-7$. Their findings indicate an overall increasing trend in galaxy bias, compared to their previous work \citep{Harikane_2016}. As a result of this increasing galaxy bias, there is a corresponding rise in the required host halo mass threshold, notably shifting for luminous UV galaxies towards values approximating $M_h\gtrsim 10^{11.37} M_{\odot}$ around redshift $z\sim 6$, with this shift becoming more pronounced for even brighter galaxy candidates.\\
\indent Further progress at higher redshift was precluded both by the limited sample sizes at $z\sim 8$ and by the heterogeneous nature of surveys carried out in the last ten years, which often employed a "wedding-cake" depth strategy (i.e. multiple layers of exposure times). This strategy is very efficient for characterising the luminosity function across a wide dynamic range \citep{Bouwens_2015}, but it makes the measurement of the two-point correlation function (which can be used to contrain galaxy clustering) more challenging. The two-point correlation function encodes the excess probability of finding two objects, in our case galaxies, separated by a given distance. The excess is evaluated relative to a uniform distribution of sources \citep{1993ApJ...412...64L}.\\
\indent With a heterogeneous survey including regions at different depths, special care must be placed to avoid introduction of systematic uncertainty and artificial clustering signal simply because some areas have an excess of galaxies due to deeper observations. In this paper we investigate the impact of non-homogeneous survey depth on the two point correlation function measurement and propose a method to conduct an accurate analysis by careful generation of the reference random sample for use in the two point correlation function estimator. Rather than simply generating a uniform random distribution over the geometry of the survey and taking into account a binary mask to model geometry and presence of foreground bright sources that mask faint high-redshift objects, we generate the reference random sample through a Monte Carlo simulation that places artificial sources with realistic spectral energy distributions and recovers them through the full photometric pipeline. This approach improves the two-point correlation function analysis well established techniques for accurate measurements of the galaxy luminosity function.\\ 
\indent The primary objective of this paper is to showcase a novel approach through the analysis of data obtained from the Hubble Legacy Fields \citep{2021AJ....162...47B}. This study has two main goals: firstly, to validate our methodology by comparing it with traditional two-point correlation techniques using substantial lower redshift datasets. Secondly, to extend our measurements to high redshifts ($z=8$) by combining surveys with varying depths, given the challenge of obtaining large high-redshift samples independently.\\
\indent Our novel method should be ideal for applications to upcoming catalogs from the increasing number of JWST photometric surveys that are characterising the galaxy population at $z>8$ \citep{Paris_2022,Hainline_2023}.\\ 
\indent This paper is organized as follows. In Section \ref{data selection} we describe the data selection for the LBGs candidates. Variables estimation and the theory behind the ACF is presented in Section \ref{variables estimation}. In Section \ref{Clustering analysis methods} we address the different methods to generate the points catalog needed to proceed in the measurements. Our findings and numerical results are discussed in Section \ref{clustering results}. In Section \ref{summary} we presents our summary and conclusions.\\ 
\indent In this work we assume, when relevant, the cosmological parameters determined by the  Planck Collaboration \citep{ade2014planck}: $(\Omega_{M},\Omega_{\lambda}, h, \sigma_{8})=(0.315, 0.685, 0.673, 0.828)$. Magnitudes are in the AB system \citep{1983ApJ...266..713O}.

\section{Data selection}\label{data selection}
\begin{figure*}
\begin{tabular}{cc}
    \includegraphics[width=0.3\linewidth]{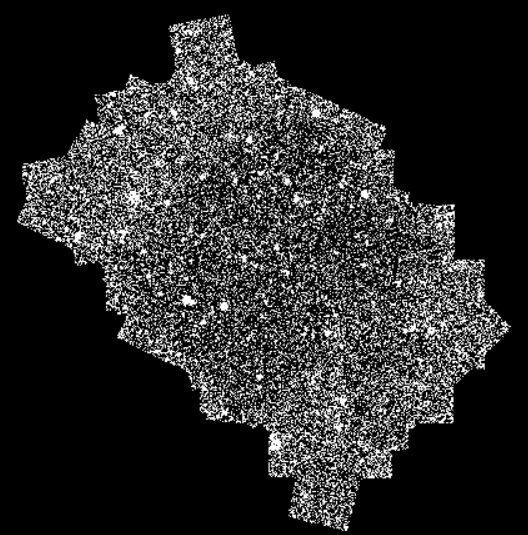}
    \includegraphics[width=0.25\linewidth]{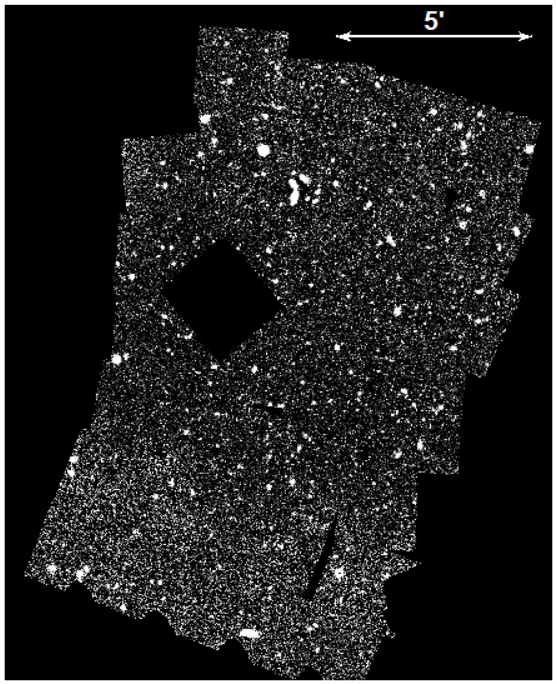}\\
    \includegraphics[width=0.35\linewidth]{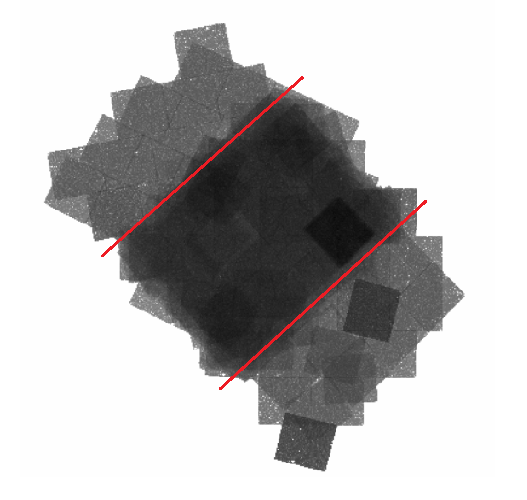}
    \includegraphics[width=0.25\linewidth]{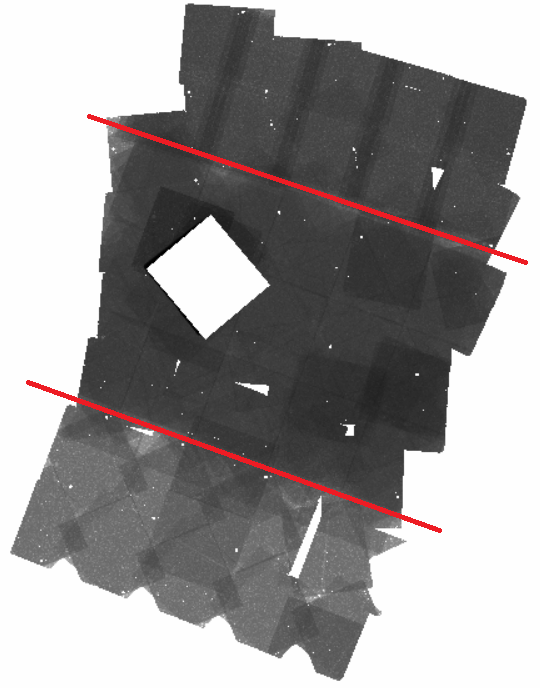}
\end{tabular}
\caption{\textbf{Left}: GOODS-North maps. \textbf{Right}: GOODS-South maps. The top row shows the science images while the bottom has rms maps in logarithmic scale and divided in sections by red lines: \textbf{Top}, \textbf{Central} and \textbf{Bottom} to show the differences in depth due a wedding-cake depth strategy of the survey (see \citealt{Grogin_2011}).}
\label{fig:GOODS_rms}
\end{figure*}

\indent This study utilises the most recent data released by the Hubble Space Telescope (HST), specifically the comprehensive collection known as the Hubble Legacy Fields\footnote{\url{https://archive.stsci.edu/prepds/hlf/}}. In our exploration of this extensively researched field, we have opted to make use of publicly available catalogs, such as the VizieR Online Data Catalog\footnote{\url{https://ui.adsabs.harvard.edu/link_gateway/2021yCat..51620047B/doi:10.26093/cds/vizier.51620047}} \citep{2021AJ....162...47B}.\\
\indent Our ultimate objective is to assemble a sizable galaxy dataset for conducting clustering analysis. To assess the influence of varying depth on source recovery and detect any potential systematic biases, we generated $\chi^2$ images by utilizing data from the $Y_{105}$, $J_{125}$, $JH_{140}$, and $H_{160}$ filters. These $\chi^2$ images will be foundamental in the development of random point catalogs, as further detailed in the subsequent sections.\\
\indent We initiated by creating a parent sample sourced from the VizieR Online Data Catalog, involving the selection of galaxies within the redshift range $z\in[4.5,8.5]$. We subsequently divided this primary sample into four redshift bins, each with a width of $\Delta z=1.0$, with the mean redshift within each bin being $\overline{z}$=5.1, 6.1, 6.8, and 7.7, respectively. To refine the analysis and identify sources in contiguous regions, we obtained the latest versions of the science and weight images from the Hubble Legacy Fields Project (\href{https://archive.stsci.edu/hlsps/hlf/}{HST Legacy field data}, \citet{bouwens_2017, whitaker_2019}). Specifically, Version 2.0 was used for the GOODS-S field, while Version 2.5 was employed for the GOODS-N field, with both images having a scale of 60 milliarcseconds per pixel.\\
\indent Subsequently, we converted the weight images into root mean square (rms) images. Considering that the GOODS-S field contains the XDF/HUDF patch, which involves much deeper observations that may introduce systematic uncertainties in the estimation of the correlation function beyond our ability to correct for them, we took an additional step to separate the XDF/HUDF region out. Furthermore, we excluded the HUDFP1 and HUDFP2 parallel ultra-deep regions within the GOODS-S field due to the absence of the required $B_{435}$ filter for $z\sim5$ selection in the HUDFP1 region, as well as the presence of multiple bright stars in the HUDFP2 region. Additionally, the selection area in the GOODS-S field was trimmed to the common overlap area in the F125W, F140W, and F160W filters to ensure a uniform selection of sources. The XDF area is not included in the rms map of GOODS-S see right panels of Fig.\ref{fig:GOODS_rms} therefore it will be analysed separately.\\
\indent The next step involved filtering the catalog by applying a magnitude cut of $M_{UV}<-19.8$ in the two GOODS regions in order to obtain a magnitude-limited (uniform) sub-sample for conducting a consistent analysis across the entire survey area and redshift range. This magnitude limit was chosen to ensure source detection above the nominal $5\sigma$ depth limit for all fields and redshifts. We decided to not apply any magnitude cut to the XDF field since the whole XDF region is internally uniform. Additionally, applying the same magnitude cut as the GOODS region would have resulted in the number of candidates that are too few to conduct a meaningful analysis. Therefore, our investigation encompasses all XDF sources, irrespective of their brightness. For reference, the average absolute magnitudes of the samples within the XDF area range from $\overline{M}_{UV}=-18.6$ to $\overline{M}_{UV}=-18.1$. Details of the data selection for each redshift bin considered in this study are summarized in Tab.\ref{tab:data selection}.\\ 
\indent For a comprehensive description of the dataset (including the determination of galaxy luminosity functions), we refer the reader to \citet{2021AJ....162...47B}.

\begin{table*}
\centering
\begin{tabular}{lccccc}
    \hline \hline
    \multicolumn{6}{c}{Number of sources}\\
    \hline
    \multicolumn{1}{c}{This work} & \multicolumn{1}{c}{Bouwens-2021} & \multicolumn{1}{c}{$\overline{z}=5.1$ } & \multicolumn{1}{c}{$\overline{z}=6.1$ } & \multicolumn{1}{c}{$\overline{z}=6.8$ } & \multicolumn{1}{c}{$\overline{z}=7.7$ }\\ 
    \hline \hline
    \textbf{XDF} & HUDF/XDF & 187 (\textbf{187}) & 146 (\textbf{146}) & 124 (\textbf{124}) & 44 (\textbf{44})\\ 
    \hline
    \multirow{2}*{\textbf{GOODS-S}} 
        & ERS & 223 (\textbf{108}) & 96 (\textbf{51}) & 69 (\textbf{36}) & 15 (\textbf{13})\\
        & C. GOODS-S & 709 (\textbf{297}) & 291 (\textbf{123}) & 129 (\textbf{50}) & 51 (\textbf{30}) \\
    \hline 
    \textbf{GOODS-N} & C. GOODS-N & 1092 (\textbf{507}) & 372 (\textbf{192}) & 259 (\textbf{150}) & 108(\textbf{73})\\
    \hline\hline
    \multicolumn{2}{c}{TOT} & 2211 (\textbf{1099}) & 905 (\textbf{512}) & 581 (\textbf{360}) & 218 (\textbf{160})\\
    \hline \hline
\end{tabular}
\caption{Summary of the LBG sources used in this work for clustering measurements in three main fields (XDF, GOODS-S and GOODS-N). The sources are extracted from \citep{2021AJ....162...47B} as discussed in Section~\ref{data selection}, and the table includes for each redshift bin the parent catalog source number and in \textbf{bold} between parentheses the objects selected for analysis.}
\label{tab:data selection}
\end{table*}
\section{Clustering estimation}\label{variables estimation}

\indent First, we derive the observed ACF indicated as $ w_{obs}(\theta)$, that measures the excess probability of finding two objects -- galaxies in our case -- at an angular separation $\theta$. We use the ACF estimator defined by \citet{1993ApJ...412...64L}:
\begin{equation}\label{estimator}
    w_{obs}(\theta)= \frac{DD(\theta) - 2DR(\theta) + RR(\theta)}{RR(\theta)},
\end{equation}
where $DD(\theta)$ is the number of pairs of observed galaxies with angular separations $(\theta \pm \delta\theta)$. 
$RR(\theta)$ is the analogous quantity determined from a random catalog with the same geometry and selection properties as the observed catalog. $DR(\theta)$ is the number of random-galaxy cross pairs.\\
\indent For the functional form of ACF, we assume a power law parameterization $w(\theta)=A_{w}\theta^{-\beta}$, where $w(\theta)$ is our estimator. The finite size of this survey leads to an underestimation of the measured quantities, which can be rectified by incorporating a constant known as the integral constraint (IC; see \citealt{10.1093/mnras/253.2.307}):
\begin{equation}
    w_\textrm{true}(\theta)=w_\textrm{obs}(\theta) + IC
\end{equation}
estimated as:
\begin{equation}\label{IC}
\begin{split}
    IC &=\frac{1}{\Omega^{2}}\int_{1}\int_{2}w_\textrm{true}(\theta)d\Omega_{1}d\Omega_{2}\\ &=\frac{\sum_{i} RR(\theta_{i}) w_\textrm{true}(\theta)}{\sum_{i} RR(\theta_{i})} =\frac{\sum_{i} RR(\theta_{i}) A_{w}\theta^{-\beta}_{i}}{\sum_{i} RR(\theta_{i})}
\end{split}
\end{equation}
where $w_\textrm{true}(\theta)$ is the intrinsic ACF and $w_\textrm{obs}(\theta)$ is the measurement within the survey area. For consistency with previous work (e.g.,\citealt{Lee_2006,Overzier_2006,Barone_2014}), we fix $\beta = 0.6$ and use linear binning with $12".5$ width with the maximum separation at $\theta=250"$.\\
\indent When the correlation slope $\beta$ is fixed, the quantity $IC/A_{w}$ only depends on the size and the shape of the survey area and can be  determined from Eq.\ref{IC}. The dimensionless parameter $A_{w}$ is the only variable that needs to be fit as:
\begin{equation}\label{estimator model}
    w_\textrm{model}(\theta)=A_{w}(\theta^{-\beta} - IC/A_{w})
\end{equation}
by maximizing the likelihood expressed as:
\begin{equation}\label{likelihood}
    \mathcal{L}= \prod_{i\in\textrm{fields}}\frac{1}{\sigma_{obs}(\theta)_{i}\sqrt{2\pi}} e^{-\frac{1}{2}\left(\frac{w_{obs}(\theta)_i-w_m(\theta)_i}{\sigma_{obs}(\theta)_{i}}\right)^2}
\end{equation}
where $w_{obs}(\theta)_i$ and $w_m(\theta)_i$ represent the observed and modelled ACF measurements while $\sigma_{obs}(\theta)_{i}$ the uncertainties in field $i$. Errors in the ACF, for both fields, are estimated using bootstrap resampling \citep{Ling_1986}.\\
\indent To estimate the galaxy bias, we approximate the real-space correlation function with a power law 
$\xi(r)=(r/r_{0})^{-\gamma}$ with $\beta\equiv\gamma-1$, where the coefficients are related to the dimensionless $A_{w}$ coefficient by the Limber transform \citep{adelberger_2005}:
\begin{equation}\label{r0}
    A_{w}=\frac{r_{0}^{\gamma}B[1/2,1/2(\gamma -1)]\int_{0}^{\infty}dzN(z)^{2}f^{1-\gamma}g(z)^{-1}}{[\int_{0}^{\infty}dzN(z)]^{2}}
\end{equation}
Here $f$ is the transverse comoving distance, i.e. $f\equiv(1 + z)D_{A}(\theta)$ where $D_A$ the angular diameter distance. $N(z)$ is the redshift distribution of the dropouts that takes into account efficiency and completeness of source identification, $B(t_1,t_2)= \Gamma(t_1)\Gamma(t_2)/\Gamma(t_1+t_2)$ is the beta function and $g(z)\equiv c/H(z)$ is the comoving distance.\\
\indent After finding $A_w$ by maximizing the likelihood Eq.\ref{likelihood} we are able to invert Eq.\ref{r0} in order to obtain the associated correlation length $r_{0}$. From the real-space correlation function $\xi(r)$ we define the galaxy bias $b$ through the ratio between the galaxy variance $\sigma_{8,g}$ and the linear matter fluctuation $\sigma_{8}(z)$ both in a sphere of comoving $8h^{-1}cMpc$ radius:
\begin{equation}
    b=\frac{\sigma_{8,g}}{\sigma_{8}(z)}
\end{equation}
assuming $\sigma_{8}(0)=0.828$ from the cosmology model and the galaxy variance expressed as:
\begin{equation}
    \sigma_{8,g}^{2}=\frac{72(r_0/8h^{-1}cMpc)^{\gamma}}{(3-\gamma)(4-\gamma)(6-\gamma)2^{\gamma}}
\end{equation}

\section{Clustering analysis methods}\label{Clustering analysis methods}
To conduct measurements of galaxy clustering, it is necessary to generate a catalog of random points for the calculation of the observed angular correlation function (ACF) estimator $w_{obs}$, as defined in Eq.\ref{estimator}. Previous studies adopted a simpler approach when generating the random points catalogs, mainly because they were based on surveys with relatively uniform depth.\\
\indent In such cases, the primary concern is modeling the survey geometry and accounting for obstructions caused by bright foreground sources. However, the GOODS-CANDELS survey exhibits non-uniform depth, as shown in Fig.\ref{fig:GOODS_rms}. Consequently, in order to evaluate the pair count $RR(\theta)$ for the random points, we investigate the impact of the random catalog generation on the ACF measurement. This investigation is motivated by the observations presented in Fig.\ref{fig:GOODS_rms}, where the survey fields consist of sub-fields with varying individual depths. Moreover, distinct patterns are discernible in small regions, characterized by non-uniform depths (e.g., refer to the bottom right image illustrating the RMS map of GOODS-South in its bottom and top regions). All these factors contribute to the heterogeneity of the survey.\\
\indent Since the survey's completeness depends on the depth of observations, there may be an excess or lack of sources in the observed catalog at specific locations and preferred angular separations associated with the patterns in the RMS maps. In this section, we introduce two approaches for clustering analysis in the presence of non-uniform-depth data and evaluate their respective strengths. Our analysis focuses on candidate galaxies selected through photometry in the CANDELS data over GOODS at an average redshift of $z=5.1$ \citep{leethochawalit2022quantitative}, in both cases the random points catalog will be of dimension $N_r=20N_d$ where $N_d$ is the number of data in the analysis at a certain redshift value.\\
\indent Specifically, we explore two different aspects that are fundamental for galaxy clustering measurements, the first one being the generation of the random points catalog. The second one is robustness of the analysis with respect to the use of different regions of the survey.

\subsection{Random point catalog generation}\label{Random point catalog generation}

\subsubsection{Uniform random points}\label{Uniform random points}

This represent the standard approach of generating random point by sampling a uniform distribution throughout the area covered by the survey, in our case the GOODS field.\\
\indent To take into account the survey geometry (and presence of bright foreground sources and/or regions with missing data), the uniform random catalog is filtered through a binary map (reject or accept) constructed considering all pixels in the image that have a well-behaved RMS map value and are not part of the segmentation maps of (bright) sources. In practice, for faint and relatively rare galaxies at high redshift, one can use the full segmentation map without noticeable difference \citep{Barone_2014}.

\subsubsection{Random points from source recovery map}\label{Random points from source recovery map}

This second method, which is introduced in this paper, aims to incorporate the varying depth of observations, including subtle features arising from mosaicing and dithering, as observed in the RMS map depicted in Fig. \ref{fig:GOODS_rms}. In order to improve the fidelity of our source detection and completeness modeling in the random catalog, we qualitatively account for these data characteristics by implementing an artificial source injection and recovery process within the region covered by the ACF observations. The procedure utilizes the publicly available source injection and recovery code, {\tt GLACiAR2} \citep{leethochawalit2022quantitative}, and employs a hit-and-miss Monte Carlo technique, assuming a model luminosity function for the LBG population. This code is typically employed to compute completeness in determining the luminosity function and, as such, realistically simulates all pertinent factors that influence the likelihood of identifying galaxies in observations, including non-uniform depth and photometric scatter in the Lyman break and/or photometric redshift selection algorithm.\\
\indent For the GOODS field, we injected a total of $N_g=500$ galaxies within each $M_{UV}$ and redshift bin. These galaxies possess Sérsic light profiles with $n=1$, random inclinations and ellipticities, and a physical size of $r=1.25$ kpc at $z=5$. At other redshifts, the size of the injected galaxies was scaled as $1/(1+z)$. They exhibit a power-law intrinsic spectral energy distribution (SED) with a Lyman break, where the power-law UV slope is randomly drawn from a Gaussian distribution with a mean of $\beta_{\rm{UV}}=-2.2$ and a standard deviation of $\sigma=0.4$. The magnitudes of the galaxies in each photometric band were determined based on their input spectral energy distributions. These artificial sources were then randomly injected into the science image using a uniform distribution of positions. The process of source extraction followed the same steps utilized for the real data sample. The magnitudes, redshifts, and positions of the recovered galaxies were recorded and employed for subsequent analysis.\\
\indent Since the catalog of recovered artificial sources was generated assuming a flat input luminosity function $\Phi(M)$, we performed a Monte Carlo hit-and-miss procedure to select the final sample of random points for clustering analysis. The probability function for selection was defined as:
\begin{equation}
p(M) = \frac{\Phi(M)}{\Phi(M_{\text{lim}})}
\end{equation}
where $\Phi^*=10^{-2.81}$, $M^*=-20.68$, and $\alpha=-1.58$ were used for the Schechter luminosity function \citep{leethochawalit2022quantitative}.\\

\subsection{Spatial dependence of the clustering signal and random points methods comparison}\label{Spatial dependence of the clustering signal and random points methods comparison}

\begin{figure*}
\begin{tabular}{cc}
\includegraphics[width=0.4\linewidth]{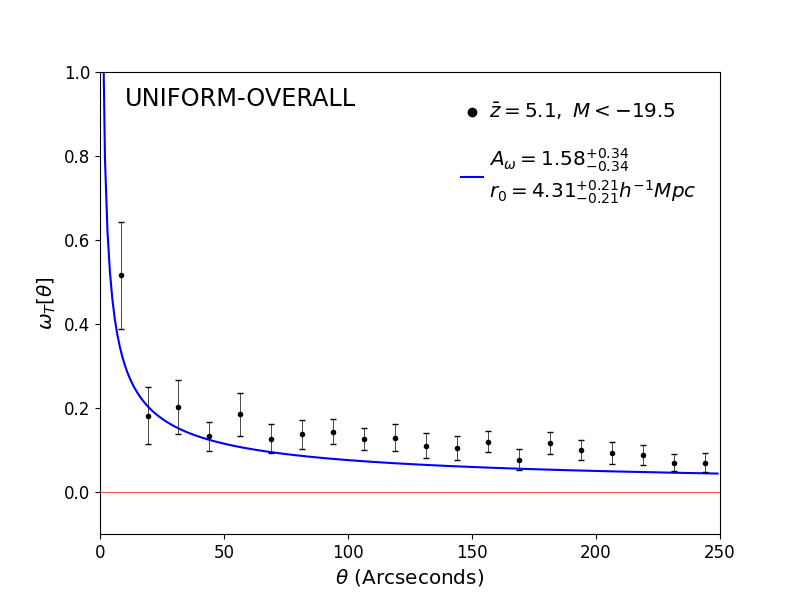} \includegraphics[width=0.4\linewidth]{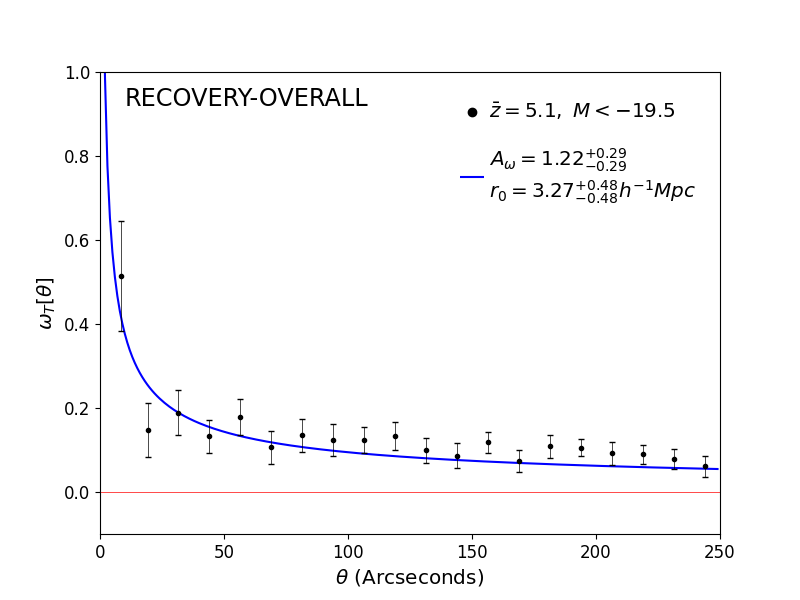}\\
\includegraphics[width=0.4\linewidth]{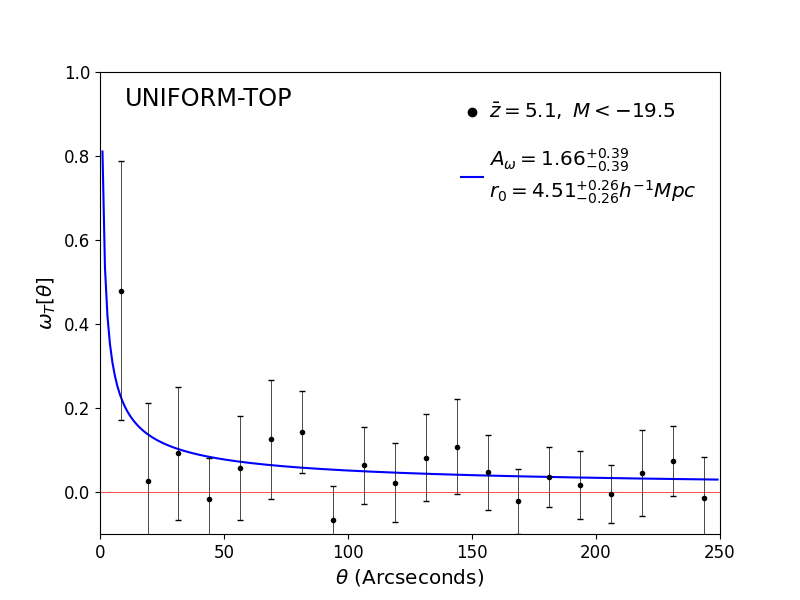} \includegraphics[width=0.4\linewidth]{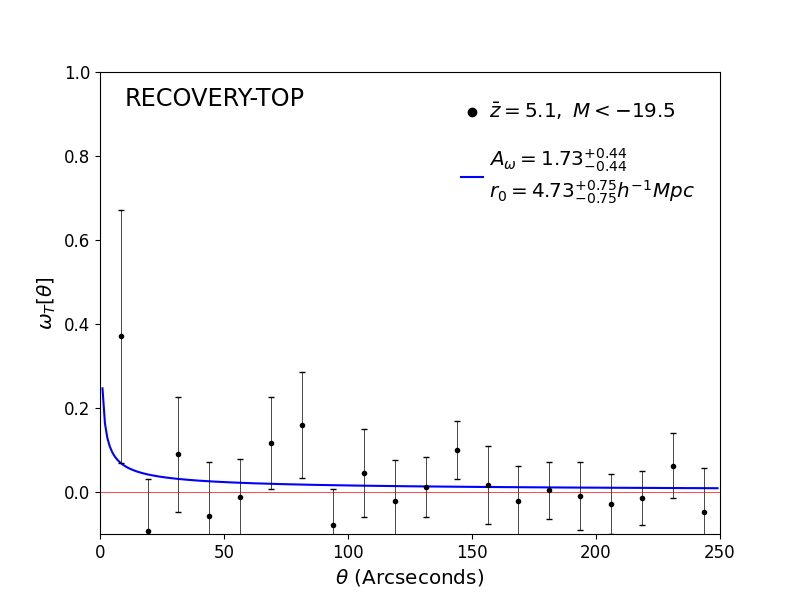}\\
\includegraphics[width=0.4\linewidth]{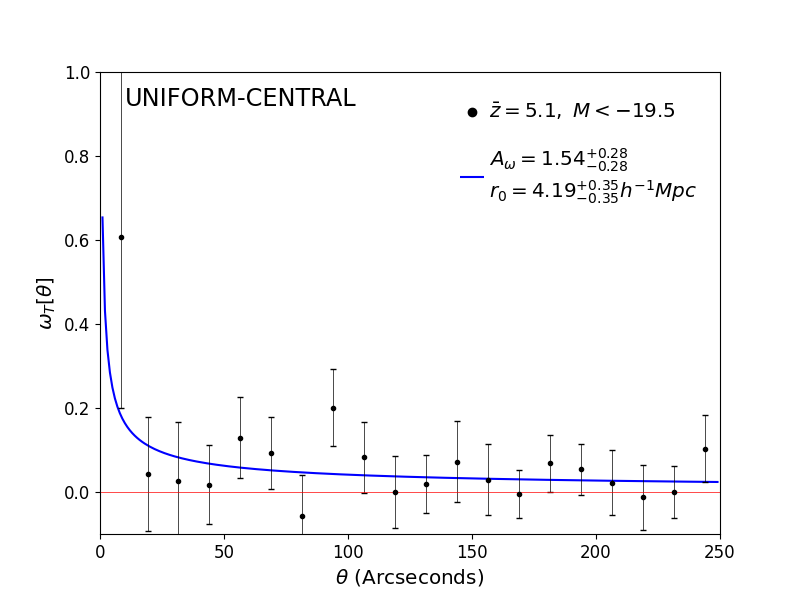} \includegraphics[width=0.4\linewidth]{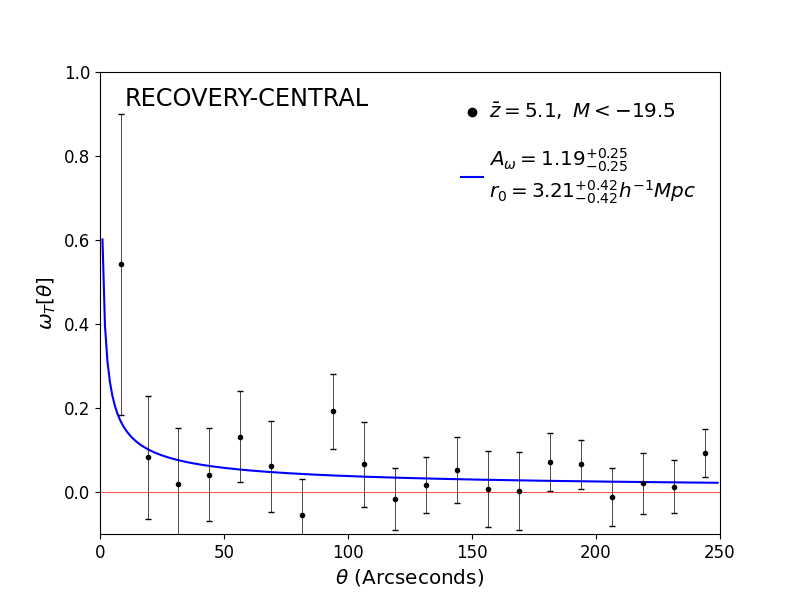}\\
\includegraphics[width=0.4\linewidth]{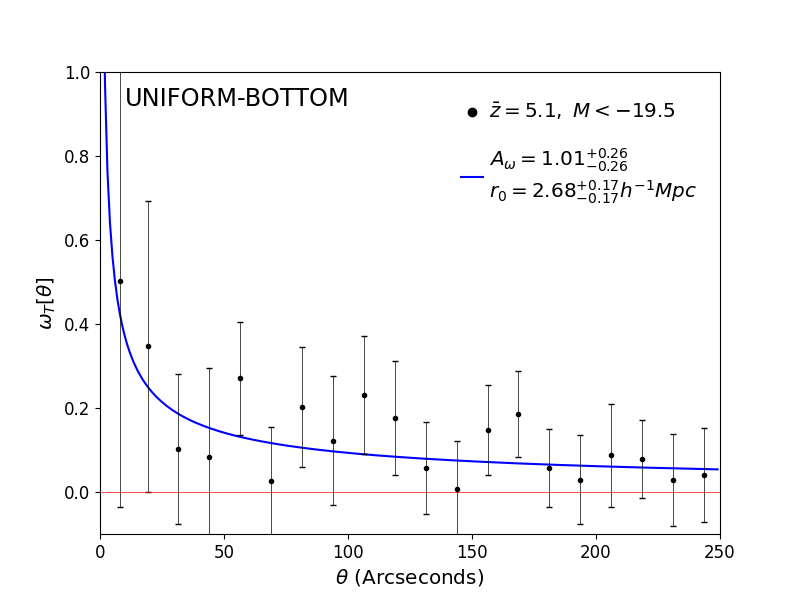} \includegraphics[width=0.4\linewidth]{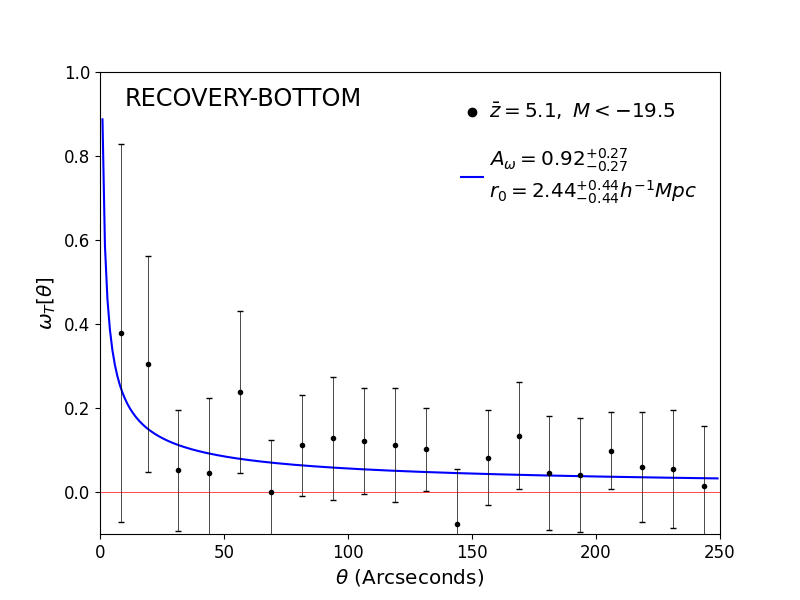}
\end{tabular}
\caption{ACFs for LBGs at $\overline{z}=5.1$ with $M_{UV}<-19.5$. On the left column we can see results obtained using a catalog of random points generated by a uniform distribution Sec.\ref{Uniform random points}, while on the right the analysis conducted with a random point recovery catalog Sec.\ref{Random points from source recovery map}. From top to bottom we can see the results produced using data from: \textbf{Overall}, \textbf{Top}, \textbf{Central} and \textbf{Bottom} region of the survey in Fig.\ref{fig:GOODS_rms}. From this figure we can see that the central region is the one that better represent the analysis on the overall survey, this is because is the most deep as we can see in the bottom row of Fig.\ref{fig:GOODS_rms}.}
\label{fig:ALL REGIONS}
\end{figure*}

Surveys covering a wide area may show variations in depth, resulting in a layered structure resembling a `wedding cake', as illustrated in Fig.\ref{fig:GOODS_rms} for the CANDELS survey. Specifically, the GOODS fields consist of three distinct regions, each characterised by an approximately constant exposure time (i.e., map RMS values).\\
\indent To study the impact of non-uniform exposure, we perform clustering analysis using both methods to generate the random catalog, as described in the previous sections, on the three separate regions: upper, middle and lower. The visual results are presented side by side in Fig.\ref{fig:ALL REGIONS}.\\
\indent The top row of Fig.\ref{fig:ALL REGIONS} illustrates the results obtained from analyzing the complete data sample using both random point catalog generation methods. It is evident that utilizing a uniform random point distribution in the autocorrelation function (ACF) measurement leads to an overestimation when compared to the best-fit power law.\\
\indent This behavior becomes apparent in the top-left corner of Fig.\ref{fig:ALL REGIONS} and manifests at large angular separations ($\theta\sim100"$). It arises due to the introduction of artificial clustering signals induced by non-uniform depth variations across the survey area, with fluctuations of the order of one WFC3 field of view (i.e., $\gtrsim 100"$). These variations can be visually correlated with the typical fluctuations in the RMS maps shown in Fig.\ref{fig:GOODS_rms}. By employing a uniformly generated random point catalog, these variations are not taken into account.\\
\indent In contrast, this issue is alleviated by adopting our proposed improved procedure for generating the random point catalog using a Monte Carlo recovery process (top-right of Fig.\ref{fig:ALL REGIONS}), although it does not entirely eliminate the artificial clustering signals.\\
\indent Interestingly, employing the latter approach yields a larger uncertainty in the correlation length $r_0$ compared to the uniform random catalog counterpart. The difference amounts to a factor of two, and we consider this estimation to be more representative of the true errors in the measurement.\\
\indent Examining the analysis conducted on the individual sub-regions, we observe a consistent trend in terms of error estimations for $r_0$. In all cases, the estimations tend to be $1.5-2\sigma$ larger compared to the results obtained from the uniformly generated sample.\\
\indent By dividing the field into sub-regions of approximately uniform depth, we avoid significant overestimation of the ACF. This outcome arises from the fact that considering spatially confined regions leads to reduced variations in depth among the individual pointings superimposed in the mosaic. However, each region yields ACF and $r_0$ estimations that differ from one another at a level surpassing the nominal associated error.\\

\section{Clustering Results}\label{clustering results}
\subsection{Angular correlation function and Bias}\label{acf and bias}
\begin{figure*}
\begin{tabular}{cc}
    \includegraphics[width=0.4\linewidth]{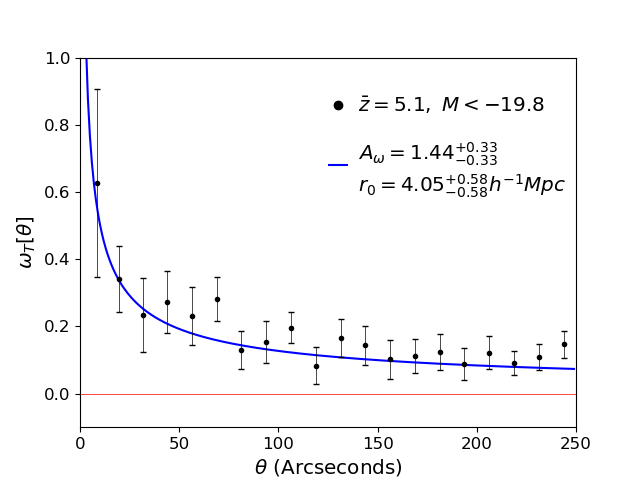}
    \includegraphics[width=0.4\linewidth]{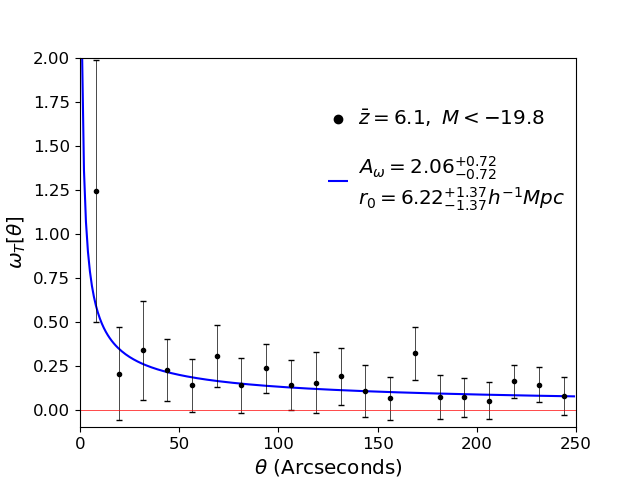}\\
    \includegraphics[width=0.4\linewidth]{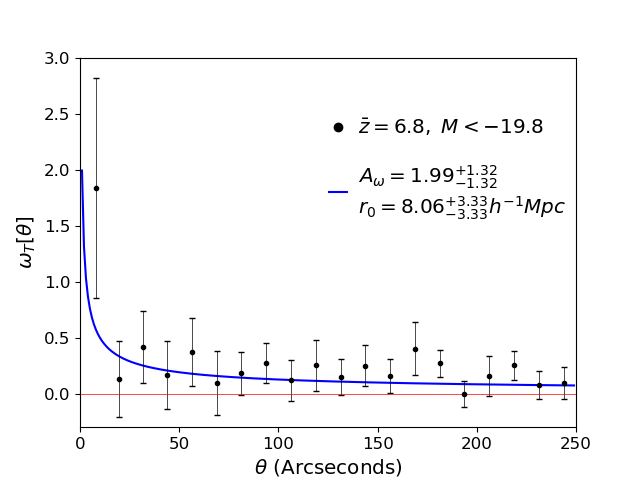}
    \includegraphics[width=0.4\linewidth]{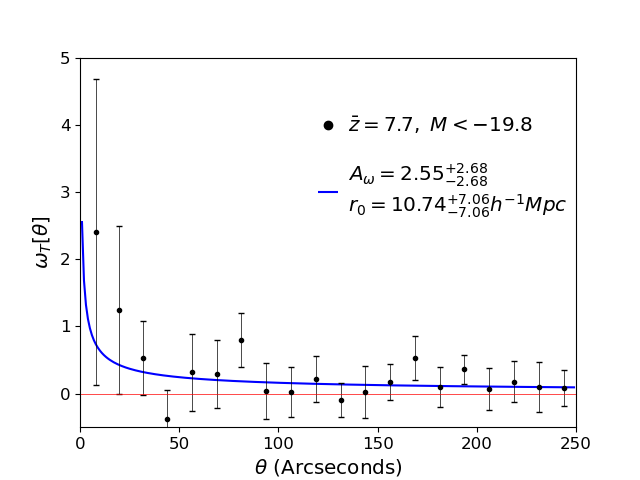}
\end{tabular}
\caption{ACFs of LBGs in the GOODS catalog at four different redshifts. The results presented were obtained with random points generated from a recovery process Section \ref{Random points from source recovery map} and all the galaxies contained in the GOODS-N and GOODS-S region. The parameter $A_w$ is a dimensionless coefficient obtained by fitting the ACF estimator Eq.\ref{estimator}, is linked by Eq.\ref{r0} to the correlation length $r_0$. Both these parameters present an increasing trend with redshift as excepted from theory and previous findings on this topic, we can see how the power law model well represent the ACF measurements.}
\label{fig:GOODS all magcut}
\end{figure*}

\begin{figure*}
\begin{tabular}{cc}
    \includegraphics[width=0.4\linewidth]{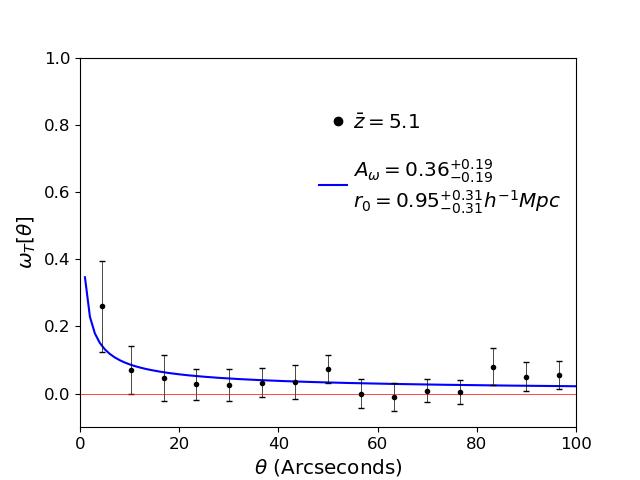}
    \includegraphics[width=0.4\linewidth]{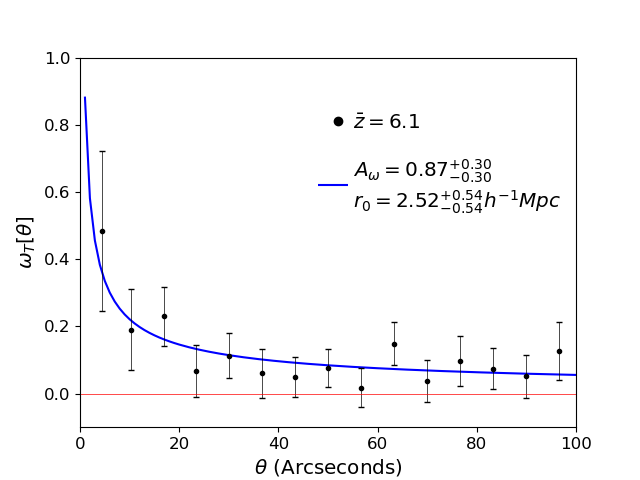}\\
    \includegraphics[width=0.4\linewidth]{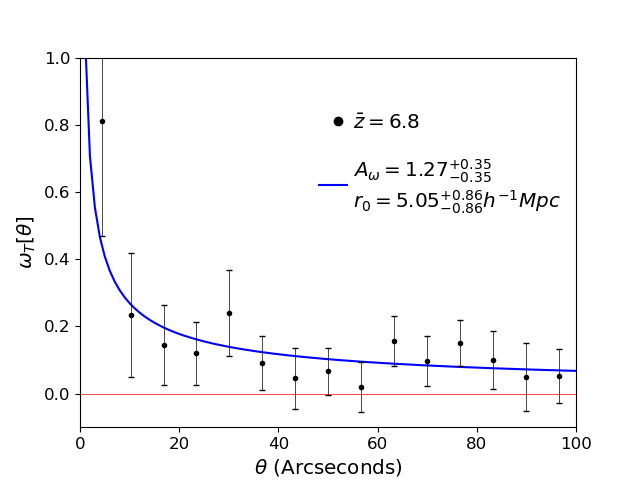}
    \includegraphics[width=0.4\linewidth]{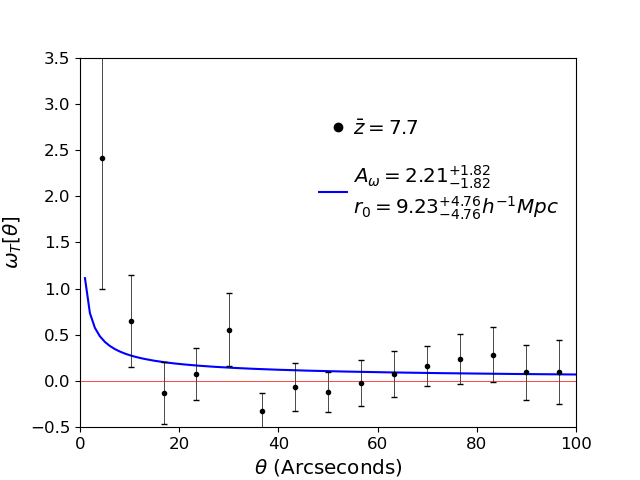}
\end{tabular}
\caption{\textbf{Top}: ACF for LBGs in GOODS-S deep field XDF catalog. The results presented were obtained with random points generated from a recovery process Section \ref{Random points from source recovery map} and the entire set of data from Tab.\ref{tab:data selection} without applying a cut in magnitude. The parameter $A_w$ is a dimensionless coefficient obtained by fitting the ACF estimator Eq.\ref{estimator}, is linked by Eq.\ref{r0} to the correlation length $r_0$. In this figure we can see the same increasing trend of the clustering parameters as observed in Fig.\ref{fig:GOODS all magcut}, values are lower because of the smaller area of XDF and the reduced pool of available candidates.}
\label{fig:XDF all magcut}
\end{figure*}

\begin{figure*}
\begin{tabular}{cc}
\includegraphics[width=0.4\linewidth]{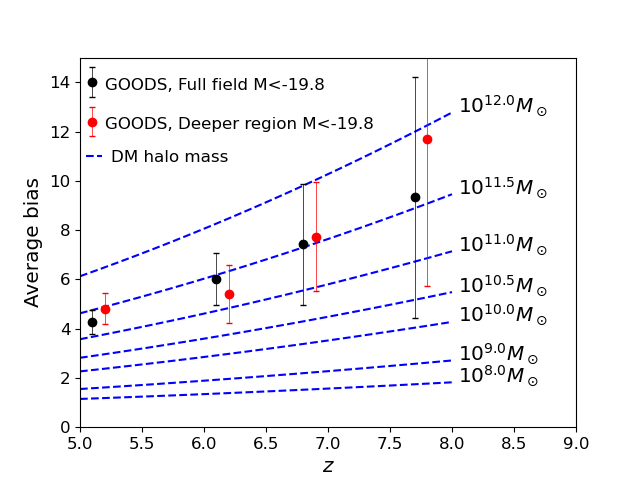}
\includegraphics[width=0.4\linewidth]{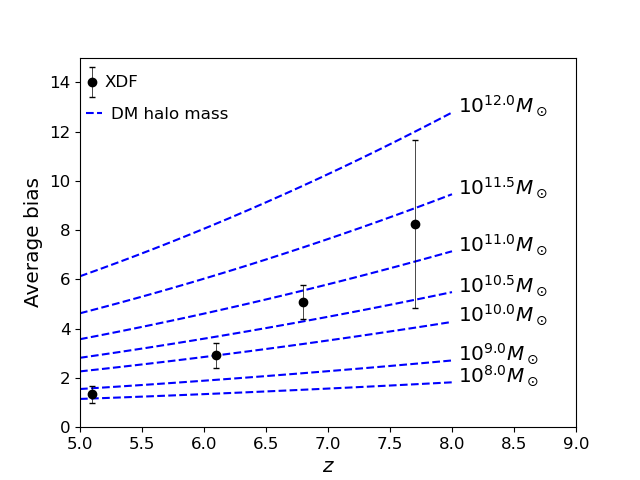}
\end{tabular}
\caption{Bias evolution as a function of redshift from 5.1 $\leq$ z $\leq$ 7.7. Dotted lines represents the dark matter halo bias from the \citet{10.1046/j.1365-8711.1999.02692.x} mass function. On the left panel the red dots are slightly shifted for better visualisation.}
\label{fig:b(z)}
\end{figure*}

\begin{figure}
\begin{tabular}{c}
    \includegraphics[width=.9\linewidth]{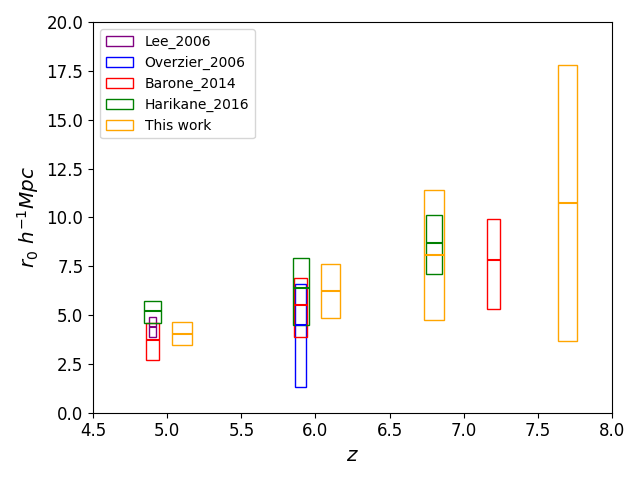}
\end{tabular}
\caption{Comparison of the correlation length $r_0$ with previous studies conducted on a sample of UV bright LBGs at similar redshifts \citep{Lee_2006,Overzier_2006,Barone_2014,Harikane_2016}. The height of the boxes represents the corresponding error bars, while the thick horizontal line within each box represents the peak measurement of the respective study. \textbf{Note}: the width of the boxes does not convey any physical information; it is solely used for enhanced visualization purposes.}
\label{fig:r0 comparison}
\end{figure}

\begin{figure*}
\begin{tabular}{cc}
    \includegraphics[width=0.4\linewidth]{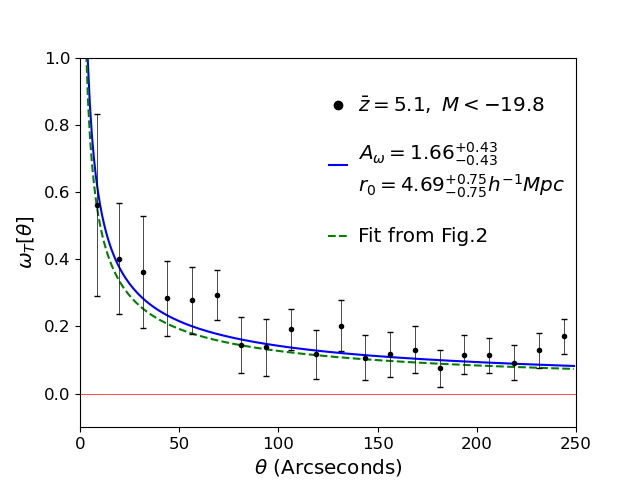}
    \includegraphics[width=0.4\linewidth]{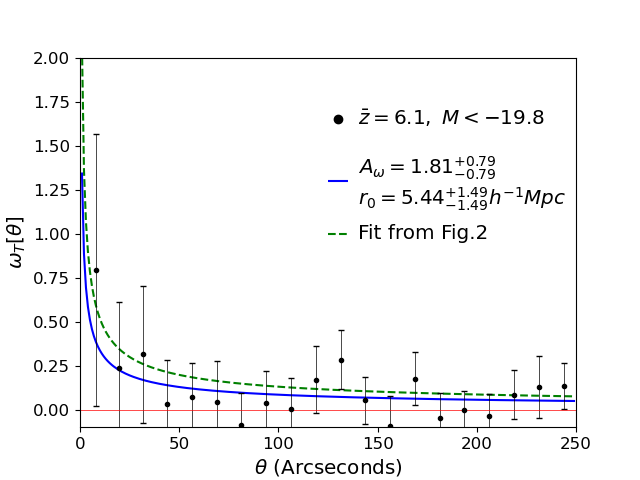}\\
    \includegraphics[width=0.4\linewidth]{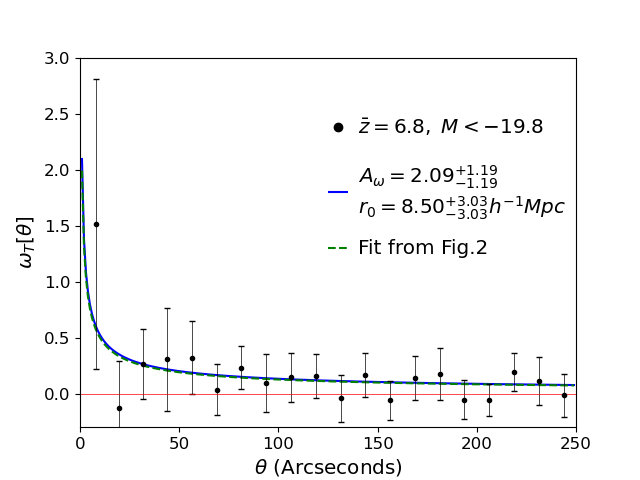}
    \includegraphics[width=0.4\linewidth]{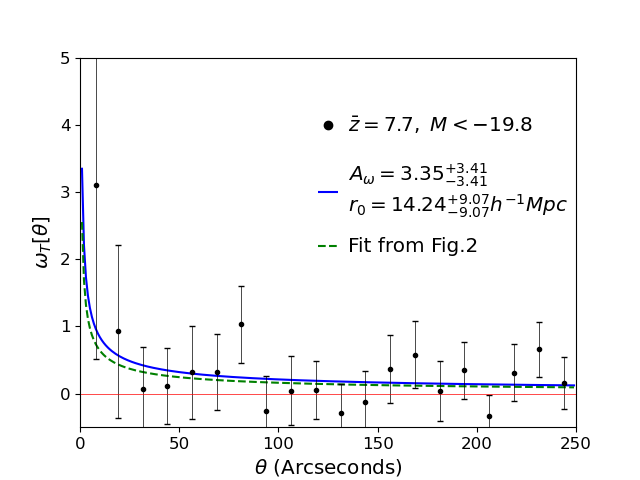}
\end{tabular}
\caption{ACF for LBGs in GOODS catalog. The results presented were obtained with random points generated from a recovery process Section \ref{Random points from source recovery map} and all the galaxies contained in the GOODS-N and GOODS-S central region (bottom row Fig.\ref{fig:GOODS_rms}). The blue fitting line return the dimensionless parameter $A_w$ linked by Eq.\ref{r0} to the correlation length $r_0$ , while the green dashed line represent the fit obtained by considering the entire combined GOODS catalog.}
\label{fig:GOODS cen magcut}
\end{figure*}

\indent In Fig.\ref{fig:GOODS all magcut} we present the ACF estimated from the candidate galaxies in the combined GOODS North and GOODS South catalog, while Fig.\ref{fig:XDF all magcut} displays the ACF estimated from the XDF region. In both figures, we provide the measurements and associated uncertainties for each separation bin, along with the best-fitting model (indicated by the blue line) and its associated parameters $A_w$ and $r_0$ reported as sub-plot legends. For comparison we present the results for all the redshift bins considered in the analysis $\overline{z}=5.1$, $\overline{z}=6.1$, $\overline{z}=6.8$ and $\overline{z}=7.7$ in separate panels.\\
\indent As discussed in the data selection process, Sec.\ref{data selection}, we are accounting for a cut in magnitude in both GOODS-N and GOODS-S considering only candidates with $M_{UV}<-19.8$, while for XDF we did not consider any constraint in the magnitude. This means that the average absolute magnitudes of the samples are lower for this last field, ranging from $\overline{M}_{UV}=-18.6$ to  $\overline{M}_{UV}=-18.1$ (without any apparent trend in redshift). The results of the ACF analysis for all fields are summarized in Tab.~\ref{tab:all results}, where the left part reports the parameters obtained from the combined GOODS calatog and the right the results from the analysis on XDF.\\
\indent First, we find that a power-law fit presents an overall good description of the data at all redshifts considered, including for our highest redshift sample at $\overline{z}=7.7$ in both Fig.\ref{fig:GOODS all magcut} and Fig.\ref{fig:XDF all magcut} (albeit the uncertainty is larger). From lower to higher redshift, if we consider a fixed angular scale, the ACF is increasing. As a result the two clustering parameters $A_w$ and $r_0$ are increasing accordingly. Qualitatively, the XDF angular correlation function shape and redshift evolutionary trends are similar, but the increase in clustering strength is more pronounced for this small-area field.\\
\indent Fig.\ref{fig:b(z)} shows the evolution of the galaxy bias, obtained at the four different redshifts studied from the ACF, overlaid on dashed curves that correspond to the bias of dark-matter halos at constant mass (derived following \citealt{10.1046/j.1365-8711.1999.02692.x}). The comparison shows that the galaxy bias increases with redshift tracing approximately constant halo mass $M_{h}\sim 10^{11.4}~\mathrm{M_{\odot}}$ for our large-area magnitude-limited GOODS sample. The results we obtain are fully consistent with theory (e.g., \citealt{10.1046/j.1365-8711.1999.02692.x}) and with previous observational studies (e.g.,\citealt{Lee_2006,Overzier_2006,Barone_2014,Harikane_2016, Hatfield_2018, Qiu_2018,Harikane_2022}).\\ 
\indent At a redshift of $z\sim6$, our investigation of a luminous sample of LBGs unveils a galaxy bias of $b=6.0\pm1.1$, which closely resembles the outcomes reported by other groups in the same redhisft regime (e.g.,\citealt{Harikane_2016,Hatfield_2018}). In their study, they determine a minimum halo mass of $M_h\sim10^{11.5}M_{\odot}$ using a simple abundance model, consistent with the predictions made in this work at the same redshift with respect to the halo mass function depicted in the left panel of Fig.\ref{fig:b(z)}. Furthermore, \citet{Qiu_2018} presents the galaxy bias obtained at a redshift of $z\sim7$ through clustering measurements with a similar sample. The resulting value is $b=8.6^{+2.2}_{-2.8}$, which is consistent within the measurement uncertainties with the value found in our study at an average redshift of $z=6.8$.\\ 
\indent In the recent investigation conducted by \citet{Harikane_2022}, they provide a summary of their findings based on the angular correlation function of over $N=4\cdot10^6$ galaxy candidates at $z\sim2-7$. The far-right panel of the figure presents the results obtained for luminous ultraviolet (UV) galaxies at a redshift of $z=5.9$. According to their analysis, the halo mass of these galaxies is estimated to be approximately $M_h\sim10^{11.05}M_{\odot}$, with an uncertainty consistent with the findings of \citet{Barone_2014} at a similar redshift of $z\sim6$.\\ 
\indent From the right panel of Fig.~\ref{fig:b(z)}, we see that the XDF field shows a smaller bias compared to GOODS at all redshifts. This is expected because the sources are on average fainter as the data is deeper and no magnitude cut is applied to retain a sufficient number of galaxies for clustering analysis. However, in this case there is a strong redshift evolution in the inferred characteristic halo mass of the population, which at face value seems to evolve from $M_h\sim 10^8 M_{\odot}$ at $z=5.1$ to $M_h\sim 10^{11.5} M_{\odot}$ at $z=7.7$. Given the small area of the field, we consider this trend most likely to originate from cosmic variance in the clustering measurement itself, which is not captured by the error bars in the figure (covering only statistical uncertainty). Incidentally this trend was also present in the XDF analysis of \citet{Barone_2014} and we have here extended it again from $z\lesssim 7$ to $z\lesssim 8$.\\
\indent To provide a visual comparison with previous investigations on the topic, we present in Fig.\ref{fig:r0 comparison} the comparison in between the correlation lengths $r_0$. Through a comprehensive analysis of low redshift data ($z\lesssim 7.0$), our work exhibits a similar trend to that presented in other independent studies conducted on a comparable sample of LBGs that are bright. We find that the characteristic correlation length increases as we move towards higher redshifts. Our findings for each redshift bin align with previous redshift measurements reported in the literature (e.g.,\citealt{adelberger_2005,Lee_2006,Barone_2014,Harikane_2016,Hatfield_2018,Qiu_2018,Harikane_2022}), with consistency observed within the 1$\sigma$ uncertainty range.\\
\indent An important novelty of our analysis is the extension of this trend to higher redshifts ($z=7.7$), specifically towards the mid-point of the epoch of reionization. We identify a correlation length of $r_0=10.74^{+7.06}_{-7.06}$ $h^{-1}Mpc$, which aligns with the aforementioned trend. However, it is worth noting that the large error bars are primarily a consequence of the limited quantity of LBGs available in our sample at such high redshift. In the coming years, this limitation is expected to improve significantly thanks to the utilization of the James Webb Space Telescope (JWST) in new surveys (e.g.,\citealt{Paris_2022, Hainline_2023}).\\
\indent As demonstrated in Sec.\ref{Spatial dependence of the clustering signal and random points methods comparison}, our analysis reveals that the clustering parameters are not significantly overestimated when studying similar-depth sub-regions. To ensure consistency, we extend our investigation to include the entire field while focusing on measurements obtained solely from the central region of both the GOODS-N and GOODS-S catalogs (see Fig.\ref{fig:GOODS_rms}). In Fig.\ref{fig:GOODS cen magcut}, we present the results, wherein the blue fitting curve represents the central region data, and we overlay the best-fit model obtained using the complete sample of galaxies labeled in green.\\
\indent Tabulated in central column of Tab.\ref{tab:all results} are the results for both $A_w$ and $r_0$, along with the associated galaxy bias, for all redshifts considered. Comparison of these measurements reveals consistency within the typical uncertainties of $1\sigma$, which can also be visually observed in the left panel of Fig.\ref{fig:b(z)}. It is worth noting that the galaxy bias $b$, when derived solely from the central region, marginally exceeds the value obtained from the overall analysis. This slight difference can be attributed to the fact that the central region represents the deepest portion among the three sub-regions, thus effectively capturing the general behavior exhibited by the entire field.

\begin{table*}
\centering
\begin{tabular}{cccccccccc}
    \hline \hline
    \multicolumn{1}{c}{} & \multicolumn{3}{c}{GOODS overall region} & \multicolumn{3}{c}{GOODS central region} & \multicolumn{3}{c}{XDF}  \\
    \hline
    $\overline{z}$ & $A_{w}$ & $r_{0}$ & $b$ & $A_{w}$ & $r_{0}$ & $b$ & $A_{w}$ & $r_{0}$ & $b$ \\
    \hline \hline
    $5.1$ & $1.44^{+0.33}_{-0.33}$ & $4.05^{+0.58}_{-0.58}$ & $4.27^{+0.49}_{-0.49}$ & $1.66^{+0.43}_{-0.43}$ & $4.69^{+0.75}_{-0.75}$ & $4.81^{+0.62}_{-0.62}$ & $0.36^{+0.19}_{-0.19}$ & $0.95^{+0.31}_{-0.31}$ & $1.33^{+0.35}_{-0.35}$ \\[.2cm]
    $6.1$ & $2.06^{+0.72}_{-0.72}$ & $6.22^{+1.37}_{-1.37}$ & $6.03^{+1.06}_{-1.06}$ & $1.81^{+0.79}_{-0.79}$ & $5.44^{+1.49}_{-1.49}$ & $5.42^{+1.18}_{-1.18}$& $0.87^{+0.30}_{-0.30}$ & $2.52^{+0.54}_{-0.54}$ & $2.92^{+0.50}_{-0.50}$ \\[.2cm]
    $6.8$ & $1.99^{+1.32}_{-1.32}$ & $8.06^{+3.33}_{-3.33}$ & $7.42^{+2.45}_{-2.45}$ & $2.09^{+1.19}_{-1.19}$ & $8.50^{+3.03}_{-3.03}$ &  $7.73^{+2.21}_{-2.21}$ & $1.27^{+0.35}_{-0.35}$ & $5.05^{+0.86}_{-0.86}$ & $5.10^{+0.69}_{-0.69}$ \\[.2cm]
    $7.7$ & $2.55^{+2.68}_{-2.68}$ & $10.74^{+7.06}_{-7.06}$ & $9.33^{+4.90}_{-4.90}$ & $3.35^{+3.41}_{3.41}$ & $14.24^{+9.07}_{-9.07}$ & $11.69^{+5.96}_{-5.96}$& $2.21^{+1.82}_{-1.82}$ & $9.23^{+4.76}_{-4.76}$ & $8.26^{+3.41}_{-3.41}$ \\
    \hline \hline
\end{tabular}
\caption{Galaxy clustering parameters obtained from ACF analysis with a fixed $\beta=0.6$. For the combined GOODS analysis only the bright LBGs are considered, with $M_{UV}$<-19.8, while for XDF we have not applied a magnitude cut for lack in candidates, see Tab.\ref{tab:data selection}. \textbf{Note}: the correlation length $r_0$ is expressed in $h^{-1}Mpc$ units.}
\label{tab:all results}
\end{table*}

\subsection{Duty Cycle}\label{duty cycle}
\begin{figure}
\begin{tabular}{c}
    \includegraphics[width=.9\linewidth]{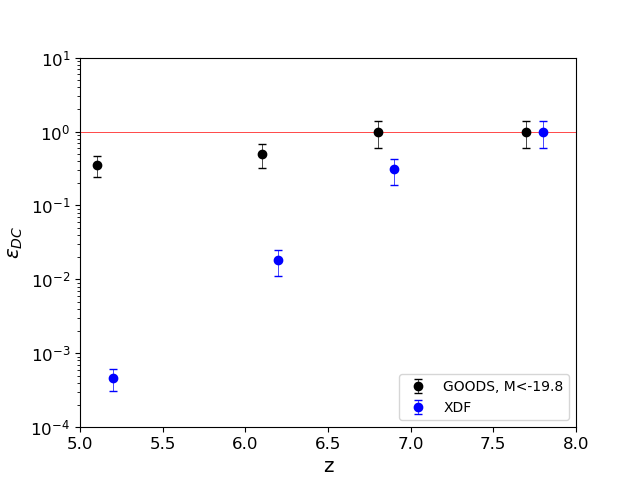}
\end{tabular}
\caption{Duty cycle $\epsilon_{DC}$ analysis conducted on the two different samples at the four redshift considered in this work. For both $\overline{z}=6.8$ and $\overline{z}=7.7$ we found the upper limit represented by $\epsilon_{DC}=1.0$.\\
The results from XDF, in blue, are slightly shifted in redshift to the right for clarity.}
\label{fig:duty_z}
\end{figure}

\indent To study star formation efficiency, we need to link the dark matter halo mass function and the luminosity function. We can do this by introducing the duty cycle ($\epsilon_{DC}$), the fraction of dark
matter halos hosting UV-bright LBGs. We can estimate the duty cycle by
matching objects at the same comoving density \citep{10.1111/j.1365-2966.2004.08059.x}.\\
\indent This procedure is conducted by matching the number of galaxies with luminosity greater than a given $L_g$ to the number of halos with mass greater than a minimum dark matter halo mass $M_h$ assuming only a galaxy in each dark-matter halo \citep{martinez_2002}:
\begin{equation}
    \epsilon_{CD}\int_{M_h}^{+\infty}n(M_h,z)dM_h = \int_{L_g}^{+\infty}\Phi(L,z)dL
\end{equation}
where $\Phi(L,z)$ if the LF at redshift $z$ and $n(M_h,z)$ is the halo MF \citep{10.1046/j.1365-8711.1999.02692.x}. By taking the intersection between the bias inferred from abundance matching, using a  Schechter LF: $\Phi^*=10^{-0.44}$,$M^*=-20.19$ and $\alpha=-1.81$ at $\overline{z}=7.7$ \citep{Bouwens_2015}, and from clustering analysis, we derive the value of $\epsilon_{DC}$. In Fig.\ref{fig:duty_z} we present the duty cycle $\epsilon_{DC}$ evolution with the redshift.\\ 
\indent We can see that at low redshift the duty cycle is contained below the unit value, while at $z>6.5$ the duty cycle tends towards unity ($\epsilon=1$), meaning that all the dark matter halos are hosting UV-bright LBGs with a mean magnitude of $M_{UV}=-19.8$. We note that the duty cycle estimated from the XDF field favours lower values of $\epsilon$ compared to the broader GOODS fields. This is a reflection of the increase in galaxy bias presented in the right-hand panel of Fig.~\ref{fig:b(z)}.\\ 
\indent The values found in this work are consistent with those of the XDF analysis conducted in \citet{Barone_2014}, they present the bias measured for XDF, HUDF92 and HUDF91 range from $b\sim2$ to $b\sim4$ in between $z\sim4$ and $z\sim6$. These three results do not present any particular pattern, but show very large variations in individual measurements, supporting our interpretation of sample variance being the most likely explanation of the XDF bias and duty cycle measurements.

\section{Summary and perspectives}\label{summary}
\indent The primary goal of this paper was to extend the investigation of galaxy clustering up to a redshift of $z\sim8$. To achieve the necessary critical number of galaxies for such a study, we combined multiple surveys of varying depth and introduced a novel method to address the systematic bias in correlation measurements within regions of varying depth. To achieve our goal, we utilized the most recent data from the Hubble Space Telescope (HST) Legacy Fields, focusing specifically on three distinct fields: GOODS-N, GOODS-S, and XDF. Our analysis covered a range of mean redshifts, namely $z=5.1$, $z=6.1$, $z=6.8$, and $z=7.7$. By exploring these fields comprehensively, we aimed to provide an exhaustive analysis that surpasses previous studies in this field.\\ 
Here, we summarize the main aspects addressed and results obtained in this work:\\
\indent $\bullet$ In order to gain a better understanding of various approaches to the analysis of the angular correlation function (ACF), we conducted a preliminary study. This involved considering the different depths of the catalogs used and incorporating random points generated through a recovery process, as described in Sec.\ref{Clustering analysis methods}. The purpose of this investigation was to explore the impact of these factors on the ACF analysis and assess their implications for our research.\\
\indent $\bullet$ To apply our innovative ACF estimation procedure, we utilized a combined GOODS catalog as well as the XDF field. The corresponding ACF measurements are presented in Fig.\ref{fig:GOODS all magcut} for the combined GOODS catalog and Fig.\ref{fig:XDF all magcut} for the XDF field. At a mean redshift of $z=7.7$, we obtained novel measurements of the galaxy bias. For the combined GOODS catalog, the galaxy bias was determined to be $b=9.33^{+4.90}_{-4.90}$, while for the XDF field, the galaxy bias was found to be $b=8.26^{+3.41}_{-3.41}$. These measurements represent important contributions to our understanding of galaxy clustering at this high redshift.\\
\indent $\bullet$ In order to validate our results, we conducted a thorough comparison with previous studies on the topic, and we found agreement between our findings and those from the literature, as demonstrated in Fig.\ref{fig:r0 comparison}. To further strengthen our analysis, we performed a supplementary validation by focusing on the central deep region of the GOODS catalog. Our results from this validation analysis were consistent with our main study, providing additional support for the robustness and reliability of our method. This validation analysis is presented in Fig.\ref{fig:GOODS cen magcut}. Together, these comparisons and supplementary analyses reinforce the credibility and significance of our study.\\
\indent $\bullet$ In Fig.\ref{fig:b(z)}, we have summarized all the ACF results and depicted them graphically alongside the trend of dark matter halo bias from the model by \citet{10.1046/j.1365-8711.1999.02692.x}, considering different mass values. Notably, all three results at $z=7.7$ consistently suggest a dark matter halo mass in the range of $M_h=10^{11.4\sim11.5}M_\odot$. This finding indicates that as we move towards higher redshift, LBGs located within larger dark matter halos exhibit higher UV luminosity. Moreover, these results are consistent with previous studies on high-redshift galaxies (e.g.,\citealt{adelberger_2005, Lee_2006, Barone_2014, Harikane_2016, Hatfield_2018, Qiu_2018, Harikane_2022}). Our findings suggest that galaxies with greater mass and luminosity display higher levels of clustering compared to previous studies at similar redshifts. This implies that these galaxies are hosted by more massive dark matter halos, providing valuable insights into the relationship between galaxy properties and their underlying dark matter structures.\\
\indent $\bullet$ We investigate the duty cycle of galaxies within dark matter halos using an abundance matching framework. We confirm the results 
from a previous study \citet{Barone_2014}, where our measured duty cycle approaches unity above $z>6.5$ (see Fig.\ref{fig:duty_z}). Furthermore, we extended the duty cycle trend to $z\sim 8$ with our novel determination of the ACF at $z=7.7$.\\
\indent The extension of the clustering analysis to a redshift of $z=7.7$ was conducted using a sample size of $N_g=160$ objects derived from surveys with varying depths. This serves as a reference point for future forecasts of galaxy clustering studies at redshifts $z\gtrsim8$ using the James Webb Space Telescope (JWST). With the current preliminary determinations of galaxy luminosity functions (e.g.,\citealt{leethochawalit2022quantitative, donnan_2023}) and ongoing photometric surveys (e.g.,\citealt{Paris_2022, Hainline_2023}), it is anticipated that high-quality photometric samples will become available within the next 12-24 months. These advancements will enable the study of the connection between galaxies and their associated dark matter halos during the earliest stages of cosmic reionization. This prospect holds great promise for furthering our understanding of the formation and evolution of galaxies in the early Universe.

\section*{Acknowledgements}
We thank the anonymous referee for useful suggestions and
comments that have improved the manuscript. This research was supported by the Australian Research Council Centre of Excellence for All Sky Astrophysics in 3 Dimensions (ASTRO 3D), through project number CE170100013. We thank Kristan Boyett for help in editing the manuscript for language and clarity.

\section*{Data Availability}

The data used to conduct the analysis presented in this paper are available in "VizieR Online Data Catalog" \citep{2021AJ....162...47B}.



\bibliographystyle{mnras}
\bibliography{example} 



\appendix

\bsp	
\label{lastpage}
\end{document}